\documentstyle[12pt]{article}\textheight 240 mm\textwidth 160 mm
\newcommand{\be}{\begin{eqnarray}}
\newcommand{\ee}{\end{eqnarray}}
\newcommand{\Z}{{\bf Z}}
\newcommand{\R}{{\bf R}}

\newcommand{\cH}{{\cal H}}

\newcommand{\ov}{\overline}

\newcommand{\sss}{\scriptstyle}

\newcommand{\Chi}{\mbox{\Large $\chi$}}

\newcommand{\Th}{\mbox{\Large $\Theta$}}

\newcommand{\NP}[1]{Nucl.Phys.\ {\bf #1}\ }
\newcommand{\PL}[1]{Phys.Lett.\ {\bf #1}\ }
\newcommand{\CMP}[1]{Comm.Math.Phys.\ {\bf #1}\ }

\begin{document}
\noindent G\"{o}teborg ITP 92-50\\
\noindent Nov 1992\\
\vspace*{10 mm}
\begin{center}
{ \large INTERACTION AND MODULAR INVARIANCE OF STRINGS \\ON CURVED
MANIFOLDS\footnote{Talk presented by S.H. at the 16'th Johns Hopkins' Workshop,
G\"oteborg, Sweden, June 8-10, 1992}\\\vspace*{10 mm}}
 Stephen Hwang
\rm and  Patrick Roberts\\
\rm Institute of Theoretical Physics\\
S-412 96 G\"{o}teborg, Sweden\\
\vspace*{10 mm}
\normalsize
\end{center}
\begin{abstract}
 We review and present new results for a string moving on an $SU(1,1)$ group
 manifold. We discuss two classes of
 theories which use discrete representations. For these theories the
representations forbidden by unitarity
decouple and, in addition, one can construct modular invariant partition
functions. The partion functions do, however,
 contain divergencies due to the time-like direction of the $SU(1,1)$
 manifold. The two classes of theories have the corresponding central
 charges  $c=9,6,5,9/2,\ldots$ and $c=9,15,21,27,\ldots$ Subtracting two
 from the latter series of central charges we get the Gervais-Neveu series
 $c-2=7,13,19,25$. This suggests a relationship between the $SU(1,1)$ string
 and the Liouville theory, similar to the one found in the $c=1$ string.
 Modular invariance is also demonstrated for the principal continous
 representations. Furthermore, we present new results for the Euclidean
 coset $SU(1,1)/U(1)$. The same two classes of theories will be possible
 here and will have central charges $c=8,5,4,\dots$ and
 $c=8,14,20,26,\ldots$, where the latter class includes
 the critical 2d black hole. The partition functions for the coset theory are
 convergent.
\end{abstract}
\setcounter{page}{0}
\newpage
\pagestyle{plain}
\normalsize \noindent
\setcounter{equation}{0}
\section{Introduction}

\ \ \ \  In string theory one usually considers string backgrounds of the
following type $M^{(d)} \otimes I^{d^\prime}$, where $M^{(d)}$ is a flat
$d$-dimensional Minkowski space and $I^{d^\prime}$ an internal
$d^\prime$-dimensional compact space. This latter space is representated
by some unitary conformal field theory. Much work has been devoted to a
classification of these and although
an enormous class of possibilities have been found, a complete classification
is still lacking. The type of spaces described above, however, is not the
most general background one may think of. A more general class of theories
is obtained if we also allow ourselves to replace the flat Minkowski space
by a more general non-compact space. We consider, therefore, a class of
backgrounds of the following general form:
${\cal{M}}^{(d)}  \otimes I^{d^\prime}$. Here  ${\cal{M}}^{(d)}$ is a
non-compact $d$ dimensional space. A natural restriction on
${\cal{M}}^{(d)}$ is that it should have at most one "time" direction. Such
a restriction may not be neccessary, but it is clearly the most obvious
generalization of the flat Minkowski case. A simple counting argument
indicates that for theories with only conformal symmetry, unitarity will
allow at most one time direction.

The motivations for studying these more general backgrounds are several. I
will only mention a few. Firstly, we do not really know if flat space is a
stable background. It may be that dynamically it will flow to a fixed point,
which is some non-trivial curved space. Such a possibility can not be ruled
out, especially since we aim to describe $d$-dimensional quantum gravity. A
further motivation is that these new theories will, as it seems, be truly
low-dimensional strings, {\it i.e.} for which the total dimension of
space-time is less than 26 or 10. This also means that we would be able to
construct new unitary conformal field theories if we could factor
out all non-unitary states in a consistent fashion. Such theories may be
"non-rational" and contain an infinite number of primary fields with respect
to some extended symmetry algebra. Finally, on quite general grounds, it will
lead to a deeper understanding of string theory. Since our past experience of
string theory is to a large extent based on having a flat Minkowski space,
going beyond this case will sharpen our intuition as to why string theory seems
to be consistent.
Let me give one example. In formulating the string perturbation theory, we are
used to thinking of this in terms of Euclidean world-sheet surfaces, where the
order of perturbation is given by the genus of the surface.
This was most elegantly exploited by Polyakov in his path-integral approach
\cite{POLYAKOV}, but was already known in the first era of string theory. We
know, however,
that the string, as analyzed in the first quantized form, is really defined on
a Minkowski world-sheet. The transition from the Minkowskian to the Euclidean
signature is seldomly, if ever,
commented upon and the validity of the transition seems never to be questioned.
An issue which is, at first sight, not related to this transition, is the
corresponding rotation in space-time. In computing loops, we need to regularize
the amplitudes due to the Minkowski time.
The standard procedure is to make a Wick rotation. The transitions from
Minkowskian to Euclidean signature for the world-sheet and space-time
geometries are usually treated independently, but in fact, they are related.
The direction of the Wick rotation in space-time is connected, by the
requirement of convergent loop amplitudes, with the direction of the
corresponding rotation on the world-sheet. The question is now, for more
general space-time geometries including cases with compact time, if we can
still do Wick rotations on the world-sheet as well as in the space-time and
if so, what is the connection between the two. This question is of importance
in dealing with the divergencies encountered in the work I will present here.

In constructing conformally invariant string theories corresponding to
propagation on curved backgrounds ${\cal{M}}^{(d)}$, a large class of them may
be constructed using
Wess-Zumino-Novikov-Witten (WZNW) models. These should then be based on
non-compact Lie-groups. The simplest of all these is the group $SU(1,1)$,
which has a Minkowski signature. This model was first discussed in
ref.\cite{BAL}. Generalizations to higher dimensions using WZNW theories
have been given in \cite{BARS}-\cite{GINSPARG}. $SU(1,1)$ is a three
dimensional group and, therefore, not a completely realistic example.
It may instead be regarded as a toy model in which important properties of
string theories with "curved time" could be studied. I am strongly convinced
that in solving this simple model we will solve most of the problems
encounterd for more realistic theories. The versatility of this theory as a
toy model is further emphasized by the interpretation due to Witten \cite{W} as
a two-dimensional black hole. The connection with the Liouville theory
discussed in this work and further analyzed by many others {\it e.g.} in
refs.\cite{DVV2}-\cite{BERSHADSKY},
is a further motivation to study the $SU(1,1)$ string. In fact, this was my
original incentive to study this particular model in \cite{H}. The connection
with the Liouville theory was suggested by the free field formulation of
$SU(1,1)$,
since it used a "Liouville like" free field \cite{CUR}-\cite{DIS}, in which the
background charge corresponded to
the "strong-coupling" regime of $2d$-gravity.

Generally, the construction of consistent string theories requires at least
the following properties:

i. Unitarity of physical states\\
\indent ii. Modular invariance\\
\indent iii. Decoupling of non-unitary states in string amplitudes\\
\indent iv. Renormalizability and anomaly freedom

\noindent I will discuss the first three of these properties in this talk.
The fourth issue is yet unsolved, although we have studied it to some
extent. I will only give some brief comments later in this connection. Let
me start by summarizing the results below for the case of the $SU(1,1)$
string.

\noindent{\it \underline {i. Unitarity}}

The unitarity of the $SU(1,1)$ string was studied in \cite{H} for the
bosonic case and in \cite{HH} for the $N=1$ fermionic case. By unitarity we
here mean the question of whether the physical states, {\it i.e.} the states
which satisfy the usual Virasoro conditions have non-negative norm. The results
of our investigations were that for conformal anomalies $c>3$, the theory was
indeed unitary for the principal continous unitary representations of
$SU(1,1)$ and for the discrete unitary representations under the restriction
on the spin, $j>k/2$ (in our conventions both $j$ and $k$ are negative). An
analogous result holds for the fermionic case.

\noindent{\it \underline {ii. Modular invariance}}

In an earlier work \cite{HHRS}, we proposed a partition function, including
only the allowed discrete representations, which was modular invariant for
integer values of $k$, the Kac-Moody anomaly. This partition function
included an infinite number of new sectors of states realizing momentum
and winding states on a
non-Abelian group manifold which were necessary due to the special topology
of $SU(1,1)$. The sectors of states were named "non-Abelian"
winding sectors, although stricly speaking, a better
terminology is Weyl translation sectors. We have extended these results and we
now have modular invariance
in the following cases:

1. $k+2$ negative integer, discrete representations\\
\indent 2. $1/(k+2)$ negative integer, discrete representations\\
\indent 3. Principal continous representations\\
\indent 4. $SU(1,1)/U(1)$ for the cases 1, 2 and 3 above.

\noindent It should be remarked here that the first three cases all involve
partition functions that contain divergencies. Consequently, the statement of
modular invariance is formal and may be invalidated by a proper
regularization. In the last case of the coset $SU(1,1)/U(1)$ the winding
sectors are absent and these
divergencies are, consequently, removed. It is modular invariant in the same
way as any unitary CFT. The conformal anomaly for the case $k$ integer is
$c=9,6,5,9/2,\ldots$ and for $1/(k+2)$ integer, $c=9,15,21,27,\ldots$. The
corresponding numbers for the coset theory are given by subtracting one from
above. In particular we have for $1/(k+2)=-4$ that $c=26$ for the coset.
This case is of particular interest, since it corresponds to the critical
Euclidean black hole. By subtracting two from above we get another interesting
series
of central charges;
$c=7,13,19,25,\ldots$. This series has been suggested \cite{Ger},
as the possible central charges for a consistent quantum Liouville theory.
This connection is probably not a coincidence, since the coset theory may be
described by a Liouville-like field coupled to a c=1 matter field. Hence, the
subtraction of one from the coset central charge is due to the matter field
and we arrive at the central charge of the Liouville-like theory.

\noindent{\it \underline {iii. Decoupling in string amplitudes}}

The question of whether the truncation of the discrete series of
representations is consistent in scattering amplitudes, {\it i.e.} that no
non-unitary states will propagate in amplitudes, is essential for a
consistent interacting string. We have been able to show, by extending an
argument due
to Gepner and Witten for the compact case \cite{GW}, that for the cases when
we have modular invariance {\it i.e.} if  $k$ or $1/(k+2)$ are integers, then
we
either have that the forbidden representations decouple or there is no
propagation at all. Thus, assuming propagation among the allowed
representations, then we have our decoupling theorem. We have also studied
the string amplitudes using a free field representation. In this representation
the fields in the
different sectors will be able to interact with each other through an
"interpolating"
field in much the same way as Ramond and Neveu-Schwarz fields can interact
in the the ordinary fermionic string. The possible interactions among the
different representations is at present under study. It appears that there will
be some differences compaired to the Clebsch-Gordon couplings
of the horizontal $SU(1,1)$. For instance, we will not be able to couple two
discrete primaries to a principal continous one.

\noindent{\it \underline {iv. Divergencies}}

In computing the partition functions we will encounter two types of
divergencies.
The first one is due to the infinte dimensional representations of $SU(1,1)$.
It is always possible to introduce a regulator to make the
infinte sums well-defined. For the discrete representations, the divergencies
will cancel in our modular invariant partition functions and we can remove the
regulator. The finite expression found in this way is the same as one would
find by a $\zeta$-function regularization. For the continous
representations, the removal of the regulator is a more delicate
problem than for the discrete case. The divergencies of this first type exist
for the full $SU(1,1)$ theory, but are absent in the Euclidean coset.

The second type of divergence is due to the time-direction of the $SU(1,1)$
theory and hence, is also not present for the Euclidean coset. It is analogous
to the divergence found for the flat case prior to Wick rotation. This
type of divergence has proven to be very difficult to deal with. It appears not
to be a
problem of a curved time, but rather of the {\it compactness} of the time
direction which leads to divergent sums instead of integrals. Although we
have studied the problem to some depth, we have not found any satisfactory
resolution. Our approach relies on starting on a Minkowskian
world-sheet\footnote{This idea is due to Bo Sundborg} and
then deriving a generalized Wick rotation. In doing this, one encounters a
modified modular transformation on a Minkowski world-sheet. The Wick rotation
derived by such an approach does not seem to give a Euclidean modular
invariance for our theory. This is, however, a preliminary result and we will
not discuss it further in this talk. One may hope that these divergencies may
be
regularized on the infinite covering space of $SU(1,1)$. We have at present no
understanding of the string theory on this space. The main unsolved problems
are the decoupling of non-unitary representaions and construction of modular
invariant partition functions.\vspace{5mm}

Let me now present and discuss some of our results a little more extensively.

\setcounter{equation}{0}
\section{The $SU(1,1)$ theory}
\ \ \ \ The $\hat{su}(1,1)$ currents $J^\pm (z)$ and $J^3 (z)$ and their
counterparts
in the other chiral sector satisfy the $\hat{su}(1,1)$ current algebra
\be
J^+(z)J^-(w)&=&{k\over (z-w)^2}+{2\over z-w}J^3(w)\nonumber\\
J^3(z)J^3(w)&=&{k/2\over (z-w)^2}\nonumber\\
J^3(z)J^\pm(w)&=&\pm {1\over z-w}J^\pm (w)\label{eq:algebra}
\ee
In addition to these currents we have the primary fields $V_{j,m}(z)$, where
$j$ is the spin corresponding to a Casimir $j(j+1)$ and $m$ is the eigenvalue
of $J^3_0$. For the discrete unitary representations of $SU(1,1)$,
$j$ belongs to the set of negative integers or half-integers. Other values are
possible for
multi-valued representations. For the principal continous representations
$j=-1/2+i\rho$, with $\rho$ being real. When acting on the $SL_2$ invariant
vacuum
these primary fields will define primary states, $\mid 0;j,m>=V_{j,m}(0)\mid
0>$.
The energy-momentum tensor is constructed by the standard procedure
\be
T(z)={1\over k+2}:\left[ (J^3(z))^2+{1\over 2}J^+(z)J^-(z)+{1\over 2}
J^-(z)J^+(z)\right]:\label{eq:emtensor}
\ee
and has a conformal anomaly $c=3k/(k+2)$. Note that if $k<-2$, which
is the case assumed here, then $c$ is greater than three, which is the physical
dimension of $SU(1,1)$.

The state space, constructed by acting with the negative modes of the
currents on the primary states, will in general contain negative norm states.
This is normal for a string theory and to project out a physical and unitary
state space, we require states to satisfy the Virasoro conditions
\be
(L_n-\delta_n)\mid {\rm phys}>=0,~~ n\geq 0.\label{eq:phys}
\ee
The unitarity of this physical subspace, for $k<-2$, is achieved for the
principal continous representations and for the discrete representations for
$k/2<j<0$ \cite{H}.
An important property of the state space is that for certain values of $j$
and $k$ for the discrete representations there exists null-states, {\it i.e.}
states which are both primary and descendent. Such states are orthogonal to
all other states and are, consequently, formally zero. For integer values of
$k$ we have a particularly simple class of such null-primaries, {\it e.g.} of
highest weight
\be
(J^+(z))^NV_{j,j}(z)=0.\label{eq:nulls}
\ee
Analogous lowest weight fields also exist. Here $j=k/2-(N-1)/2$ for
$N=1,2,\ldots$ We should here observe that these null-primaries are absent
in the range of spins, $k/2<j<0$, which is the same range required by unitarity
of the physical
space. In fact, by studying the Kac-Kazhdan determinant \cite{KK}, one can
conclude that for the range of spins $k/2<j<-k/2-1$, there are no
null-primaries at all. This is in contrast to the $\hat{su}(2)$ case, where the
unitary representations occur for spins which have null-states. In this
case we know by the analysis of Gepner and Witten \cite{GW} that these
null-primaries play an essential r\^ole in consistency of the truncation
to the
allowed representations. For $\hat{su}(1,1)$ we may modify this argument and
prove a similar decoupling theorem for $k$ being an integer. We study a
three-point
function with one primary field having a spin $j_1$ in the forbidden region.
This is sufficient for establishing the decoupling. We use
eq.(\ref{eq:nulls}) inserted into the three-point function
\be
0=\left<(J^+(z_1))^NV_{j_1,j_1}(z_1)V_{j_2,m_2^\prime}(z_2)V_{j_3,m_3^\prime}(z_3)\right>
\label{eq:threepoint}
\ee
We proceed by using the OPE of currents with primaries to eliminate
$J^+(z_1)$. Then (\ref{eq:threepoint}) will imply
\be
\left<V_{j_1,j_1}(z_1)V_{j_2,m_2}(z_2)V_{j_3,m_3}(z_3)\right>=0.
\label{eq:decoupling}
\ee
The decoupling theorem is then concluded from this equation by observing that
for $\hat{su}(1,1)$ the identity representation contains no null-primaries.
Consequently, if there exists any propagation of states in the unitary
sector, then the non-unitary sector must decouple. For $\hat{su}(2)$ the
situation is in a sense reverse, the identity representation contains
null-primaries, so that only representations which do not contain nulls of the
form (\ref{eq:nulls}) will decouple. Another important consequence of this
difference between $\hat{su}(1,1)$ and  $\hat{su}(2)$ is that
eq.(\ref{eq:nulls}) will not, in the former case, yield additional selection
rules in correlation functions. Our argument this far has been for integer
$k$. One may generalize this result for the case of $1/(k+2)$ being an
integer. This case is more complicated. For $2j=k,k-1,\ldots$, we can still use
the null-states above. These $j$-values are rational, and hence, we must
consider representations on covering spaces of $SU(1,1)$.
There are, however, always more $j$-values on a given covering than those
above.
The trick is to generate the rest of them by Weyl translations (which will be
discussed in the next section). One can show that precisely in the cases of
$1/(k+2)$ being integers will we generate all spins on the cover of order
$-1/(k+2)$. For other values of $k$ we have at
present no similar construction and, consequently, we do not know of the
consistency of the unitarity truncation.

Let us end this section by presenting the characters of $\hat{su}(1,1)$.
We require the representations to belong to the unitary sector. Since this
sector does not contain any null states, it is straightforward to compute
the characters. The main difficulty is the fact that we are dealing with
infinite dimensional representations and, therefore, we need to regularize
the traces. We define the characters by
\be
\Chi_{j,0}(\tau,\theta) = Tr\{
e^{2\pi i[(L_0 -c/24)\tau + J^3_0 \theta]}\}\label{eq:character}
\ee
The trace is here taken over the states in the respective representations.
The sum  over $J^3_0$ eigenvalues is regularized by letting $\theta$ have a
an imaginary part of the appropriate sign. The explicit form for the discrete
representations is
\be
\Chi^{\pm}_{j,0}(\tau,\theta) = e^{2\pi i(\frac{(2j+1)^2}{4(k+2)})\tau\mp 2\pi
ij\theta}
  R^{-1}(\tau,\pm\theta)
\label{eq:ndchar}
\ee
with
\be
R(\tau,\theta) = (1-e^{2\pi i\theta})e^{\pi i\tau/4}\prod_{n=1}^{\infty}
(1-e^{2\pi in\tau}e^{-2\pi i\theta})(1-e^{2\pi in\tau})
(1-e^{2\pi in\tau}e^{2\pi i\theta}).
\label{eq:denr}
\ee
For the principal continous representations it is
\be
\Chi^{a}_{j,0}(\tau,\theta) = \eta(\tau)^{-3} e^{-2\pi
i(\frac{\rho^2}{k+2})\tau}
\sum_{n=-\infty}^{\infty} e^{2\pi i(n+a)\theta},
\label{eq:ncchar}
\ee
where $a=0$ or $1/2$ for single-valued representations. We notice that the
characters in eq.(\ref{eq:ndchar})
and (\ref{eq:ncchar}) diverge for $\theta\rightarrow 0$, as expected from the
infinite dimensionality of the representations. The sum
$\Chi_{j,0}^-+\Chi_{j,0}^+$ is, however, finite in this limit. It corresponds
to regularizing $\sum_{-\infty}^j1+\sum_{-j}^\infty 1$ to the value $2j+1$. The
same finite result is found by using a $\zeta$-function regularization and it
equals the sum of the
Plancherel measures of the two representations.\footnote{We thank Brian Greene
for this remark.}
 Using the characters above
we have not found any way to construct modular invariant partition functions.
Thus, one may suspect that we lack some essential ingredient for completing
the $SU(1,1)$ string theory. This is what we will discuss in the next section.

\setcounter{equation}{0}
\section{New sectors of states}

\ \ \ \ The topology of the $SU(1,1)$ manifold is equivalent to the space
${\bf R}^2\times S^1$. This means that we have the possibility for the string
to wind around the compact direction yielding winding states. This is a
well-known situation for a free field compactified on $S^1$. In our case there
is a complication, since the circle is embedded in a non-Abelian manifold. In
\cite{HHRS} we proposed a construction of the new sectors of states,
realizing winding states for our case. Let me here explain this construction
by making comparisons with the simple Abelian case. The discussion here will
apply to the case of integer $k$.

In the $U(1)$ theory, when we compactify a boson
on a circle of radius $r$,
where $2r^2$ is an integer (which corresponds to $k$ being integer), we may
write
the spectrum of momenta as
\be
<\alpha_0>&=& p_L={a\over 2r}+2sr\nonumber\\
<\overline{\alpha}_0>&=& p_R={a\over 2r}+2\overline{s}r.
\label{eq:smomenta}
\ee
Here $a,s,\overline{s}$ are integers with $1\leq a\leq 4r^2$. With a set of
values in the form eq.(\ref{eq:smomenta}) we can introduce a unitary
transformation
\be
\alpha_n&\rightarrow&\alpha_n^{s}=\alpha_n+2sr\delta_n\nonumber\\
\overline{\alpha}_n&\rightarrow&
\overline{\alpha}_n^{\overline{s}}=\overline{\alpha}_n+2\overline{s}r\delta_n.
\label{eq:ptransform}
\ee
On the Virasoro modes this transformation yields
\be
L_n\rightarrow L_n^{(s)}=L_n-2rs\alpha_n-2r^2s^2\delta_n,\label{eq:fvirasoro}
\ee
and we have the analogous expression for the opposite chirality. It is clear
that this transformation will transform the full set of momentum states into a
new and equivalent set of momentum states and, therefore, it is a symmetry of
the theory. This symmetry is present due to the compactness of space, yielding
a discrete spectrum of momentum and winding states. We could have arrived
at this symmetry in a different way. Let us restrict the set of values
in eq.({\ref{eq:smomenta}) to a much
smaller set, namely those in eq.(\ref{eq:smomenta}) for which $s$ and
$\overline{s}$ are zero. We could then recover the symmetry by simply applying
the transformation (\ref{eq:ptransform}) and adding all the new sectors of
states corresponding
to different values of $s$ and $\overline{s}$. The unitary transformations will
generate an
orbit of $p_L$-values for each value of $a$. It is worth noticing that in
computing the partition function,
the different orbits, labelled by $a$, will in a natural way
define characters
\be
\Chi_a=\sum_{s\in {\bf Z}}Tr\{e^{2\pi i\tau L_0^{(s)}}e^{2\pi i\theta
\alpha_0^{(s)}}\}
\label{eq:fchar}
\ee
The transformations defined above may be interpreted as Weyl translations
for an Abelian group.

In the case of $\hat{su}(1,1)$ we now proceed in a similar fashion. For
integer $k$, taking into consideration the different normalization as
compared to the Abelian case, we have in place of eq.(\ref{eq:smomenta})
\be
<J^3_0>&=&{a\over 2}+2sr_1^2\nonumber\\
<\overline{J}^3_0>&=&{a\over 2}+2\overline{s}r_1^2,\label{eq:jmomenta}
\ee
with $2r_1^2=-k\in \Z$, $1\leq a\leq -2k$. In analogy with
eq.(\ref{eq:ptransform}), we define the transformations
\be
J^3_n&\rightarrow & J^{3(s)}_n =J^3_n+2sr_1^2\delta_n\nonumber\\
\overline{J}^3_n&\rightarrow & \overline{J}^{3(s)}_n
=\overline{J}^3_n+2\overline{s}r_1^2\delta_n.\label{eq:jtransform}
\ee
In order to proceed, we must solve a problem, which we did not encounter
in the Abelian case. The transformation (\ref{eq:jtransform})
must be compatible with the $\hat{su}(1,1)$ symmetry. Indeed, if it is
unitary, it will preserve the
$\hat{su}(1,1)$ algebra. In addition, it will transform primary fields into
primary fields with respect to the transformed currents. These properties
follow directly from the $\hat{su}(1,1)$ algebra and the definition of
primary fields and they are enough to determine the transformation. One finds
\cite{HHRS}
\be
J^{\pm}_n &\rightarrow& J_n^{(s)\pm} = J^{\pm}_{n\mp 2s},\label{eq:transj}
\ee
and for the primary fields
\be
V_{j,m}(z)\rightarrow  V_{j,m}^{(s)}(z)=z^{-2ms}V_{j,m}(z).\label{eq:transv}
\ee
On the Virasoro generators these transformations induce
\be
L_n \rightarrow L^{(s)}_n=L_n -2sJ^3_n -2s^2r_1^2\delta_n.
\label{eq:transl}
\ee
Here, $L_n$ are normal ordered with respect to $J^a_n$, and
$L_n^{(s)}$ with respect to the transformed $J_n^{a(s)}$. The similarity
between this equation and eq.(\ref{eq:fvirasoro}) for the Abelian case is
striking. It should be noted from eq.(\ref{eq:transj}), that the state spaces
for different values of $s$ are distinct from each other. This is in contrast
to the Abelian case, where only the momentum values and not the higher modes
are affected by the transformation. Just as for the Abelian case, the
transformations defined above are Weyl translations of the original
operators. It is clear that these transformations are not symmetries of
the original $SU(1,1)$ theory, which means that the "momentum" and "winding"
states in eq.(\ref{eq:jmomenta}) are not compatible with the state space of
this theory. It corresponds instead to the case of
restricting eq.(\ref{eq:smomenta}) to the values $s=\overline{s}=0$ in the
$U(1)$ theory. It should be remarked that even though the Weyl translations
here, and in the $U(1)$ theory, are consequences of the presence of an $S^1$,
the converse is not neccessarily true, as is the case of {\it e g } $SU(2)$.
Another example is the Euclidean coset $SU(1,1)/U(1)$. The
corresponding Virasoro modes will be invariant under Weyl translations.
This is as expected, since the compact time direction is factored out and,
consequently, there is no $S^1$ embedded. We have now completed the
construction of the $SU(1,1)$ string and we can proceed to construct
characters for the new sectors of states and combine these into modular
invariant partition functions.

\setcounter{equation}{0}
\section{Modular invariance}

\ \ \ \ We begin our study of modular invariance by remarking that, since our
string
model has a compact time component, we will run into divergent sums due to
"momentum" and "winding" states in the real time direction. In this section we
will consider the
divergent functions to be formally defined in the sense that the modular
properties may be extracted following formal manipulations. When we consider
the Euclidean coset $SU(1,1)/U(1)$, in which the time component has been
removed, the partition functions will be convergent.

The transformation defined in the previous section induces the following new
characters
\be
\Chi_{j,0}(\tau,\theta)\rightarrow \Chi_{j,s}(\tau,\theta) = Tr\{
e^{2\pi i[(L_0-2sJ_0^3+s^2k -c/24)\tau + (J^3_0-sk)\theta]}\}.
\ee
These can be easily computed from the old characters by noting that
\be
\Chi_{j,s}(\tau,\theta)=e^{2\pi i\tau s^2k}e^{-2\pi i\theta
sk}\Chi_{j,0}(\tau,\theta -2s\tau)
\ee
Then we may write for the discrete representations
\be
\Chi^{\pm}_{j,s}(\tau,\theta) =
(-1)^{2s}e^{\pm\pi i\theta}
e^{2\pi i\tau(k+2)(s\pm\frac{2j+1}{2(k+2)})^2}e^{-2\pi
i\theta(s(k+2)\pm(j+1/2))}R^{-1}(\tau,\pm\theta).
\label{eq:dchar}
\ee
The transformed continuous characters are found similarly;
\be
\Chi^{a}_{j,s}(\tau,\theta) = \eta(\tau)^{-3} e^{-2\pi
i(\frac{\rho^2}{k+2})\tau}
\sum_{n=-\infty}^{\infty}e^{2\pi
i\tau[{\frac{(n+a-sk)^2}{k}-\frac{(n+a)^2}{k}}]}e^{2\pi
i(n+a-sk)\theta}.
\ee
Using the characters above we can proceed and seek modular invariant
combinations. We first consider the simplest extension of our algebra where
we assume $k\in\Z$, and $k<-2$. In \cite{HHRS} the following extended
character was presented for the discrete representations
\be
\Chi_j^{\pm}(\tau,\theta) &=& \sum_{s\in\Z}\Chi_{j,s}^{\pm}(\tau,\theta)
\nonumber \\
&=& \mp \frac{\Th_{\mp(2j+1),k+2}(\tau,\theta)}{
\Th_{1,2}(\tau,\theta)-\Th_{-1,2}(\tau,\theta)}
\ee
where the formal theta function  is defined by
\be
\Th_{n,k}(\tau,\theta) = \sum_{m\in\Z+\frac{n}{2k}}
exp[2\pi ik(m^2\tau-m\theta)]
\label{eq:theta}
\ee
If we wish to describe a theory that enjoys time reversal symmetry, we may
write
a modular invariant partition function as
\be
Z^D(\tau,\theta) = \sum_{k/2<j,\ov{\jmath}<0}M_{j\ov{\jmath}}
(\Chi_{\ov{\jmath}}^++\Chi_{\ov{\jmath}}^-)^*
(\Chi_{j}^++\Chi_{j}^-)
\label{eq:dpart}
\ee
where the coefficients $M_{j\ov{\jmath}}$ are non-negative integers connected
to the corresponding ones for $\hat{su}(2)$ by the substitution
($k<-4$) $j,\overline{j}\rightarrow -j-1,-\overline{j}-1$. We see from
eq.(\ref{eq:dpart}) that we only include the allowed representations. In the
diagonal combination where $M_{j\ov{\jmath}}=\delta_
{j\ov{\jmath}}$, the range of $j$ is $k/2+1\leq j\leq -1$, since this
combination of characters vanishes for non-square-integrable representaion
$j=-1/2$ as well as for $j=1/2(k+1)$. Note, that the partition function is
finite in the limit $\theta\rightarrow 0$. The divergence has been removed
in forming the combination of the two discrete representations, as remarked in
section 2. The $k=-3$ partition
function vanishes identically, and one may find two extra unitary solutions
which are not symmetric under time reversal:
\be
Z^{\pm}_{-3}(\tau,\theta)=|\Chi_{j=-\frac{1}{2}}^{\pm}(\tau,\theta)|^2 +
|\Chi_{j=-1}^{\pm}(\tau,\theta)|^2
\ee
We see that here the non-square-integrable state $j=-1/2$ appears, which is
typical of partition functions of the discrete series lacking time reversal
symmetry.

The partition functions for the continuous representations with $k\in\Z$ are
quite
simple to construct because the range of the Casimir eigenvalues are not
restricted by unitarity. Summing over integral winding sectors
($j=-1/2+i\rho$), we have
\be
\Chi_{\rho}^{a}(\tau,\theta) &=& \sum_{s\in\Z}\Chi_{\rho,s}^{a}(\tau,\theta)
\nonumber \\
&=&\eta(\tau)^{-3} e^{-2\pi i\frac{\rho^2}{k+2}\tau}\sum_{n=k}^{-1}
\Th_{2(n+a),k}(\tau,0)\Th_{2(n+a),-k}(\tau,\theta)
\label{eq:cchar}
\ee
We see that the winding sectors correct for the measure of the moduli space by
soaking up the contribution of the extra two $\eta$-functions. Taking the sum
of both continuous characters $\Chi_{\rho}^{0}+\Chi_{\rho}^{1/2}$, then the
double
theta function factor formally transforms under the modular group as a modular
form which exactly cancels the contribution from two of the $\eta$-functions
(disregarding a possible modular anomaly). Thus, we need only to integrate over
$\rho$ which acts like the momentum zero mode of an uncompactified boson:
\be
Z^C(\tau,\theta) = \int_0^{\infty} d\rho\;|\Chi_{\rho}^{0}(\tau,\theta)+
\Chi_{\rho}^{1/2}(\tau,\theta)|^2.
\ee
This partition function is also finite in the limit $\theta\rightarrow 0$. The
divergence has been absorbed into the divergent sum over $s$.

The transformation in the previous section was defined for integer $k$.
One may  extend this construction to the case $k+2=p/q$, $p<0$, $q>1$, where
$p$ and $q$ are coprime integers. We must then allow spins on the $q$-th
covering of $SU(1,1)$. We have only succeeded in constructing modular
invariant partition functions, using the representations allowed by
unitarity, for the case $p=-1$. It is also consistent with the fact that
only for this case may we generalize the decoupling theorem. The characters
are of the following form
\be
\Chi_j^{(r)\pm}(\tau,\theta) &=& \sum_{s'\in
q\Z}\Chi_{j,s'}^{(r)\pm}(\tau,\theta) \nonumber \\ &=&
\Th_{2rp\pm(2j+1)q,pq}(\tau,\theta) e^{\pm\pi i\theta} R^{-1}(\tau,\pm\theta)
\ee
where the formal (divergent) theta function is defined as in
eq.(\ref{eq:theta}). In order to relate the partition functions to the
covering groups of $SU(1,1)$, we define
\be
j^{(r)} \equiv \frac{r}{q} +j,
\ee
so that we may consider $j^{(r)}$ to be the spin of a
representation on the $q$-th covering of $SU(1,1)$. In terms of this notation
we may write the modular invariant partition functions as
\be
Z^D(\tau,\theta) = \sum_{j^{(r)},\ov{\jmath}^{(r)}}
M_{j^{(r)},\ov{\jmath}^{(r)}}
(\Chi^+_{\ov{\jmath}^{(r)}}+\Chi^-_{\ov{\jmath}^{(r)}})^*
(\Chi^+_{j^{(r)}}+\Chi^-_{j^{(r)}}).
\ee
The possible matrices, $M_{j^{(r)},\ov{\jmath}^{(r)}}$, may be found using the
known classification of non-unitary $SU(2)$ modular invariants \cite{KW} and
the
relations to $SU(1,1)$ invariants given above.

Let us now turn to the coset $SU(1,1)/U(1)$, where $U(1)$ is the compact time
component. Although the question of modular invariance of these
models has so far remained elusive,\footnote{In ref.\cite{GAW} the
3$d$ Euclidean space $SL(2,{\bf C})/SU(2)$ is discussed and a modular invariant
partition function is constructed. The coset $SL(2,{\bf C})/SU(2)$ mod ${\bf
R}$
is also considered and some indications are given that this theory is
identical to the coset $SU(1,1)/U(1)$. We have been unable to verify this. The
partition functions presented in \cite{GAW} appear to be completely different
from ours.}
we are now in the position to construct
modular invariant partition functions using the methods of
\cite{GQ}-\cite{HNY}. The state space of $\hat{su}(1,1)$ may be decomposed as
\be
\cH_j^{SU(1,1)} = \bigoplus_{m}\cH_{j,m}\label{eq:md}
\ee
where $m$ is the eigenvalue of $J^3_0$ and $j$ the spin of the ground-state.
We may further decompose the states
into a direct product of the parafermionic and $U(1)$ state
\be
\cH_{j,m} = \cH^{P\!f}_{j,m}\otimes\cH^b_{m}\label{eq:nmd}
\ee
As we have mentioned above, the coset Virasoro modes are invariant under the
Weyl translations. On the other hand, the eigenvalues of $J^3_0$ are clearly
not invariant. However, from the decomposition eq.(\ref{eq:md}) one observes
that a translation $m\rightarrow m-sk$ can be absorbed into the infinite sum
over different $m$-values (which is not restricted by the lowest or
highest weights, since it refers to the total $J^3_0$-value). The characters
are, therefore, invariant under the
Weyl translations. From the explicit expression of the character $\Chi_{j,0}$
eq.(\ref{eq:ndchar}) and the decomposition eq.(\ref{eq:nmd}), one finds for
the $SU(1,1)/U(1)$ parafermion theory
\be
\Chi^{{Pf}\pm}_{j,m}= \eta(\tau)D^{(\pm)}_{j,m}(\tau).
\ee
 Here we have defined a {\em string function}
\be
D^{(\pm)}_{j,m}(\tau) = \mp\eta(\tau)^{-3}\sum_{r=0}^{\infty}(-1)^r
e^{2\pi i\tau[(k+2)(\frac{r}{2}\mp\frac{2j+1}{2(k+2)})^2-
k(\frac{r}{2}\pm\frac{m}{k})^2]}.
\ee
It is convinient to introduce $N=-(k+2)>0$, and we write $m=m^\prime-sk$.
Summing over $s$ we may construct the following modular
function,
\be
c^{2 m^\prime-1}_{2j+1}(\tau) = \sum_{s\in\Z}(D^{(+)}_{j,m^\prime+s(N+2)}(\tau)
+D^{(-)}_{j,m^\prime+s(N+2)}(\tau)).
\label{eq:string}
\ee
It can be shown \cite{KP,HNY,ACT} that this function is equivalent to the
absolutely convergent {\em Hecke indefinite modular form}
\renewcommand{\arraystretch}{.5}
\be
c^L_M(\tau) = \sum_{\begin{array}{c}
		    {\sss -|x|<y\leq |x|}\\ {\sss (x,y)\in\R} \end{array}}
{\rm sign}(x)e^{2\pi i \tau[(N+2)x^2-Ny^2]}
\ee
\renewcommand{\arraystretch}{1}
where
\be
(x,y)\; {\rm or}\;(\frac{1}{2}-x,\frac{1}{2}+y)\in
(\frac{L+1}{2(N+2)},\frac{M}{2N})+\Z^2.
\ee
The modular properties of these functions are well known \cite{KP,ACT} and a
modular invariant partition functions can be written as
\be
Z^{P\!f}_D(\tau) = |\eta(\tau)|^2\sum_{L,M,\ov{L},\ov{M}}
N_{L,M,\ov{L},\ov{M}}c^L_M(\tau)c^{\ov{L}}_{\ov{M}}(\ov{\tau})^*.
\ee
The coefficients $N_{L,M,\ov{L},\ov{M}}$ are related to the corresponding
modular invariant partition function for the coset $SU(2)/U(1)$, which for
rational values of
$N$ are given in \cite{KW}. Note,
however, that $j$ and $m$ have exchanged r\^oles with respect to the familiar
string functions of $SU(2)/U(1)$ coset models. As remarked earlier, the above
partition functions are absolutely convergent and contain only the allowed
representations for $k$ or $1/(k+2)$ being an integer.
For $1/(k+2)=-4$ the above partition functions represent the $c=26$
Euclidean black hole.

We will finally consider the principal continous representations. Again, it
is simple to repeat the steps above with the result
\be
\Chi_{j,m}^a(\tau,\theta)=\eta(\tau)^{-2}e^{2\pi
i\tau[\frac{-\rho^2}{k+2}-\frac{m^2}{k}]},
\;\;j=-\frac{1}{2}+i\rho.
\ee
Constructing modular invariant partition functions is straightforward with
these string functions, since they are of the same form as two free bosons,
one of which is compactified on a circle of radius $\sqrt{-k/2}$.

Let us end by a remark. The modular
invariant partition functions for the coset did no require the Weyl translation
sectors. Still, it is worth noticing that in deriving the partition
functions, we made a decomposition, leading to eq.(\ref{eq:string}),
which is clearly reminiscent of these sectors. In fact, if we were
to reintroduce the $U(1)$ piece and maintain modular transformation
properties, the sum over $s$ in eq.(\ref{eq:string}), will become
the sum over winding sectors.

\vspace{1cm}

{\bf Acknowledgement} This work is a continuation of a collaboration with
M\aa ns Henningson and Bo Sundborg. We have benefited greatly from many
discussions with them.


\newpage

\begin{thebibliography}{AA}
\bibitem{POLYAKOV}A.M. Polyakov, Phys. Lett. \bf 103B \rm (1981) 207.
\bibitem{BAL} J. Balog, L. O'Raifeartaigh, P. Forg\'acs and A. Wipf, \NP{B325}
(1989) 225.
\bibitem{BARS}I. Bars and D. Nemeschansky, \NP{B348} (1991) 89.
\bibitem{BARS2}I. Bars, Proceedings of the conference "Strings and symmetries
1991", World Scientific (1991).
\bibitem{GINSPARG}P. Ginsparg and F. Quevedo, {\it Strings on curved
space-times: Black holes, torsion and duality}, preprint no.
LA-UR-92-640,NEIP-92-001; hepth@xxx/9202092.
\bibitem{W}E. Witten, Phys.Rev {\bf D44} (1991) 314.
\bibitem{DVV2}  R. Dijkgraaf, E. Verlinde and H. Verlinde, \NP{B371} (1991) 269
\bibitem{MARTINEC}E.J. Martinec and S.L. Shatashvili, Nucl.Phys. {\bf B368}
(1992) 338.
\bibitem{BERSHADSKY}M. Bershadsky and D. Kutasov, Phys. Lett. \bf B266 \rm
(1991) 345.
\bibitem{H}S. Hwang, Nucl. Phys. \bf B354 \rm (1991) 100.
\bibitem{CUR}T. Curtright and C. Thorn, Phys.Rev.Lett. {\bf 48} (1982) 1309.
\bibitem{GER}J.-L. Gervais and A. Neveu, Nucl. Phys. \bf B209 \rm
(1982) 125.
\bibitem{MAR}R. Marnelius and L. Johansson, Nucl. Phys. \bf B254 \rm
(1985) 201.
\bibitem{HWA}S. Hwang and R. Marnelius,  Phys. Lett. \bf B206 \rm (1988) 205.
\bibitem{FUJ}K. Fujikawa, T.Inagaki and H. Suzuki, Phys. Lett. \bf B213 \rm
(1988) 279.
\bibitem{DAV}F. David, Mod.Phys.Lett {\bf A3} (1988) 1651.
\bibitem{DAS}S.R. Das, S. Naik and S.R. Wadia, Mod Phys.Lett. {\bf A4} (1989)
1033.
\bibitem{DIS}J. Distler and H. Kawai,  Nucl. Phys. \bf B321 \rm
(1989) 509.
\bibitem{HH}M. Henningson and S. Hwang, Phys. Lett. \bf B258 \rm (1991) 341.
\bibitem{HHRS}M. Henningson, S. Hwang, P. Roberts  and B. Sundborg,
Phys. Lett. \bf B267 \rm (1991) 350.
\bibitem{Ger} J.-L.Gervais and A. Neveu, \PL{151B} (1985) 271: \\
A. Bilal and J.-L. Gervais, \PL{B187} (1987) 39:\\
J.-L. Gervais, \PL{B243} (1990) 85; \CMP{138} (1991) 301.
\bibitem{GW}D. Gepner and E. Witten, Nucl. Phys. \bf B278 \rm
(1986) 493.
\bibitem{KK}V. G. Ka\v{c} and D.A.Kazhdan, Advances in Math. {\bf 34} (1979)
97.
\bibitem{KW} V.G. Ka\v{c} and M.Wakimoto, Proc.Natl.Acad.Sci, USA {\bf 85}
	    (1988) 4956; \\
	    S.Lu,\PL{B218} (1989) 46; \\
	    I.C.Koh and P.Sorba, \PL{B215} (1988) 723.
\bibitem{GAW}K. Gaw\c{e}dzki, {\it Non-compact conformal field theories},
Carg\`ese summer school on "New symmetry principles in quantum field theory",
1991; hepth@xxx/9110076.
\bibitem{GQ}D. Gepner and Z. Qiu, Nucl. Phys. \bf B285
[FS19] \rm (1987) 423.
\bibitem{ZF} A.B. Zamolodchikov and V.A. Fateev, Sov.Phys. JETP \bf 63 \rm
(1985)
913.
\bibitem{KP} V.G. Ka\v{c} and D.H. Peterson, Advances in Math. {\bf 53}
	    (1984) 125.
\bibitem{HNY} K. Huitu, D. Nemeschansky and S. Yankielowicz, \\
		   USC-90/010, USC Preprint (1990).
\bibitem{ACT} C. Ahn, S. Chung and S.-H. Tye, \NP{B365} (1991) 191.
\end{thebibliography}
 \end{document}